\documentclass[referee]{raa}

\usepackage{graphicx,times,natbib,amssymb,amsmath}
\bibpunct{(}{)}{;}{a}{}{,}

\usepackage[pagebackref=true]{hyperref}

\begin{document}

\title{Potential of Constraining Propagation Parameters of Galactic Cosmic Rays with the High Energy cosmic-Radiation Detection facility onboard China's Space Station}
%\subtitle{I. Place Your Subtitle Here}

\volnopage{Vol.0 (20xx) No.0, 000--000}
\setcounter{page}{1}

\author{Zhi-Hui Xu \inst{1,2}
   \and Qiang Yuan \inst{1,2}
   \and Zhi-Cheng Tang \inst{3}
   \and Xiao-Jun Bi \inst{3,4}
   }

\institute{Key Laboratory of Dark Matter and Space Astronomy, Purple Mountain Observatory, Chinese Academy of Sciences, Nanjing 210023, China; {\it yuanq@pmo.ac.cn}\\
        \and
           School of Astronomy and Space Science, University of Science and Technology of China, Hefei 230026, China\\
        \and
           Key Laboratory of Particle Astrophysics, Institute of High Energy Physics, Chinese Academy of Sciences, Beijing 100049, China\\
        \and
           University of Chinese Academy of Sciences, Beijing 100049, China\\
\vs\no
   {\small Received 20xx month day; accepted 20xx month day}}

\abstract{Precise measurements of the spectra of secondary and primary cosmic rays are crucial for
understanding the origin and propagation of those energetic particles. The High Energy cosmic-Radiation
Detection (HERD) facility onboard China’s Space Station, which is expected to operate in 2027, will 
push the direct measurements of cosmic ray fluxes precisely up to PeV energies. In this work, we
investigate the potential of HERD on studying the propagation of cosmic rays using the measurements 
of boron, carbon, and oxygen spectra. We find that, compared with the current results, the new HERD
measurements can improve the accuracy of the propagation parameters by 8\% to 40\%. The constraints 
on the injection spectra at high energies will also be improved.
\keywords{cosmic rays: theory --- propagation --- HERD}
}

\authorrunning{Z.-H. Xu et al.}

\titlerunning{Potential of Constraining Cosmic Ray Propagation with HERD}

\maketitle

\section{Introduction}
\label{sect:intro}

The origin, acceleration, and propagation of cosmic rays (CRs) remain unresolved despite 
long time studies. One of the most important approaches to solve this problem is to measure
the energy spectra of various compositions of CRs precisely. Recent precise measurements by
mainly the direct detection experiments reveal interesting features of many nuclear species, 
including the hardenings around a few hundred GV rigidity \citep{2009BRASP..73..564P,
2017ApJ...839....5Y, 2011Sci...332...69A,2015PhRvL.114q1103A,2015PhRvL.115u1101A,2017PhRvL.119y1101A, 
2020PhRvL.124u1102A,2019PhRvL.122r1102A,2020PhRvL.125y1102A,2019SciA....5.3793A,2021PhRvL.126t1102A}, 
and softenings around $O(10)$ TV \citep{2017JCAP...07..020A,2017ApJ...839....5Y,2019SciA....5.3793A,
2021PhRvL.126t1102A,2022PhRvD.105f3021A}.
For secondary nuclei which are particularly important in understanding the propagation of CRs, 
hardening features at similar energies with primary nuclei are also found \citep{2018PhRvL.120b1101A,2021PhR...894....1A,2022SciBu..67.2162D,2022PhRvL.129y1103A}.
Those progresses of measurements triggerred many theoretical studies to discuss new implications
on the origin and propagation of CRs \citep[e.g.,][]{2011ApJ...729L..13O,2011PhRvD..84d3002Y,
2012PhRvL.108h1104M,2012PhRvL.109f1101B,2012ApJ...752L..13T,2012ApJ...752...68V,2014A&A...567A..33T,
2015ApJ...815L...1T,2016ApJ...819...54G,2016ApJ...827..119C,2018PhRvD..97f3008G,2020FrPhy..1524601Y,
2020JCAP...11..027Y,2021FrPhy..1624501Y,2021ApJ...917...61K,2021ApJ...911..151M,2022ApJ...933...78M,
2022ApJ...932...37N,2023FrPhy..1844301M,2023JCAP...02..007Z}.

The energy coverage of current precise measurements are still limited (e.g., the DAMPE measurements
of B/C and B/O reach only $\sim5$ TeV/n; \citet{2022SciBu..67.2162D}). To better understand the 
propagation properties of CRs, improved measurements in a wider energy range are crucial. 
The High Energy cosmic-Radiation Detection (HERD) facility, planned to be installed in China's
Space Station around 2027, is dedicated to measuring energy spectra of various CR species up to
PeV energies \citep{2014SPIE.9144E..0XZ}. The core of the HERD detector is a three-dimensional, 
five-side active calorimeter (CALO) detector, surrounded by Fiber Trackers (FIT) for track
measurements, Plastic Scintillator Detectors (PSD) covering the trackers for charge measurements
and anti-coincidence of $\gamma$ rays, and Silicon Charge Detectors (SCD) enclosing all the above 
sub-detectors for charge measurements \citep{2023NIMPA104867970K}. A Transition Radiation Detector 
(TRD) is employed to provide energy calibration of nuclei. The geometric factor of HERD is about
$2\sim3$ m$^2$sr for charged CR detection, and can thus extend the measurements of the spectra
of major CR components to $>$PeV energies. The novel design of HERD makes it a powerful detector 
for measurements of CR nuclei and electrons/positrons, as well as $\gamma$ rays in a wide energy 
range. HERD is expected to significantly advance our understanding of the fundamental problems 
in CR physics, as well as the probing of new physics such as the nature of dark matter particles
\citep{2016APh....78...35H}.

In this work, we study the potential of constraining injection and propagation parameters of 
HERD, assuming a simple one-zone, homogeneous propagation model of CRs. We forecast the spectral 
measurements of CR boron, carbon and oxygen according to the latest designed performance of HERD
\citep{2023NIMPA104867970K}, and employ the numerical propagation model GALPROP 
\citep{1998ApJ...509..212S} (version 56) together with a Markov Chain Monte Carlo (MCMC) 
algorithm {\tt emcee}\footnote{\url{https://pypi.org/project/emcee/}} \citep{2013PASP..125..306F}
to constrain the model parameters. This paper is organized as follows. We describe 
the propagation framework and model setting in Section~\ref{sec:methods}. 
In Section~\ref{sec:datamodel}, we will present the results of our analysis, focusing on the 
comparison with current results based on existing data. Finally, we will summarize our results 
and discuss their implications in Section~\ref{sec:conclusion}.

\section{Propagation Model of Cosmic Rays}
\label{sec:methods}

The propagation of CRs in the Milky Way, which involves a number of physical processes, 
can be described by the following equation~\citep{1964ocr..book.....G,2007ARNPS..57..285S}
\begin{eqnarray}
\frac{\partial \psi}{\partial t} & = & 
\nabla\cdot(D_{xx}\nabla \psi-{\boldsymbol V_c}\psi)
+\frac{\partial}{\partial p}p^2D_{pp}\frac{\partial}
{\partial p}\frac{1}{p^2}\psi \nonumber \\
 & - & \frac{\partial}{\partial p}\left[\dot{p}\psi
-\frac{p}{3}(\nabla\cdot{\boldsymbol V_c}\psi)\right]
-\frac{\psi}{\tau_f}-\frac{\psi}{\tau_r}+q({\boldsymbol r},p)
\ , \label{prop}
\end{eqnarray}
where $\psi$ is the differential density of CRs per momentum interval, $D_{xx}$ is the 
spacial diffusion coefficient, $D_{pp}$ is the diffusion coefficient in the momentum 
space which is employed to describe the stochastic reacceleration of CRs, ${\boldsymbol V_c}$ 
is the convection velocity, $\dot{p}$ is the momentum loss rate, $\tau_r$ is the timescale 
of radioactive decay, $\tau_f$ is the timescale of fragmentation, and $q({\boldsymbol r},p)$ 
is the source function.

We assume that the spatial diffusion coefficient, $D_{xx}$, is spatially homogeneous 
and depends on the rigidity of particles ${\mathcal R}$ with a power-law form 
\begin{equation}
D_{xx}(\mathcal R)=D_0\beta^{\eta}\left(\frac{{\mathcal R}}
{{\mathcal R}_0}\right)^{\delta},\label{Dxx}
\end{equation}
where $\beta=v/c$ is the velocity of the particle in unit of speed of light, 
${\mathcal R}_0\equiv 4$ GV is a reference rigidity, $\eta$ is introduced to tune the 
velocity dependence at low energies to better fit the data, and $\delta$ is the
rigidity dependence slope which reflects the property of the interstellar turbulence \citep{1990acr..book.....B,2002cra..book.....S}.
The convective transport is described by a velocity $\boldsymbol V_c$, which is the
fluid velocity of gas containing relativistic particles. Fitting to the data shows
that a large convection velocity is disfavored to reproduce the observed low-energy
secondary-to-primary ratios such as B/C 
\citep{1998ApJ...509..212S,2017PhRvD..95h3007Y,2019SCPMA..6249511Y}. We therefore 
ignore the convection in this work.
The scattering of particles off randomly moving magnetohydrodynamic (MHD) waves results in 
stochastic acceleration of CRs, which can be described by the diffusion in the momentum space
with coefficient \citep{1994ApJ...431..705S}
\begin{equation}
D_{pp}=\frac{4p^2v_A^2}{3\delta(4-\delta^2)(4-\delta)wD_{xx}},
\label{Dpp}
\end{equation}
where $v_A$ is the Alfven speed of magnetized disturbances, $w$ is the ratio of MHD wave 
energy density to the regular magnetic field energy density and can be effectively absorbed 
into $v_A$.

The spatial distribution of CR sources is assumed to follow the distribution 
of supernova remnants or pulsars, which is parameterized as a cylindrically 
symmetric form
\begin{equation}
f(R,z)=\left(\frac{R}{R_\odot}\right)^{\alpha}\exp\left[-\frac
{\beta(R-R_\odot)}{R_\odot}\right]\,\exp\left(-\frac{|z|}{z_s}\right),
\end{equation}
where $R_\odot=8.5$ kpc is the distance from the Earth to the Galactic 
center, $z_s=0.2$ kpc is the scale width of the vertical extension of 
sources, $\alpha=1.25$ and $\beta=3.56$ describe the radial distribution
of the sources tuned based on $\gamma$ rays \citep{2011ApJ...729..106T}.
The source function is truncated at $R_{\rm max}=$20 kpc and $z_{\rm max}=z_h$, 
where $z_h$ is the half-height of the propagation halo.

We employ the cubic spline interpolation method in the 
$\log({\mathcal R})-\log(J)$ space to describe the injection spectra 
of primary CRs \citep{2016A&A...591A..94G,2018ApJ...863..119Z}. 
Specifically, we select 7 rigidity knots to cover the rigidity range of 
current measurements:
\begin{equation}
\{\log({\mathcal R}_1),...,\log({\mathcal R}_7)\}=\{2.301,\ 2.968,\ 3.634,\ 
4.301,\ 4.968,\ 5.634,\ 6.301\},
\end{equation}
where the unit of rigidity is MV. For all primary CRs, the same spectral 
shape with different abundances is assumed. Finally, the solar modulation 
which mainly affects CR spectra at low energies is applied under the 
force-field approximation \citep{1968ApJ...154.1011G}.

\section{Analysis and results}\label{sec:datamodel}

\subsection{Fitting to existing data}
We first run a fitting to existing data, which is used for comparison. The best-fit 
results of the spectra are also the basis of the prediction of HERD observations. 
The data used in this work includes the spectra of boron, carbon, and oxygen measured 
by AMS-02 during its first seven years of operation~\citep{2021PhR...894....1A}. 
The low-energy spectra of those species measured by 
ACE\footnote{\url{http://www.srl.caltech.edu/ACE/ASC/level2/lvl2DATA\_CRIS.html}} 
within the same time window of AMS-02 are extracted \citep{2019SCPMA..6249511Y}.
The Voyager-1 also measured spectra of boron, carbon, and oxygen outside the 
heliosphere \citep{2016ApJ...831...18C}, which are helpful in determining the solar 
modulation parameter. To break the degeneracy between the diffusion coefficient and 
the halo height, we use the $\rm^{10}Be/^{9}Be$ ratio measured by several experiments
\citep{1988SSRv...46..205S,1999ICRC....3...41L,1980ApJ...239L.139W,1998ApJ...501L..59C,
2001ApJ...563..768Y,2004ApJ...611..892H,2021Univ....7..183N}. 
The observational time periods of $\rm^{10}Be/^{9}Be$ differ among each other, and 
thus there are large uncertainties of their solar modulation parameters. We use the 
modulation potential retrieved from the Cosmic Ray Database 
(CRDB)\footnote{https://lpsc.in2p3.fr/crdb}~\citep{2014A&A...569A..32M,2020Univ....6..102M}, 
based on the neutron monitor data~\cite{2017AdSpR..60..833G}. The data are summarized in Table \ref{table:data}.
%and calibrating the H and He fluxes~\cite{2016A&A...591A..94G} as well as the neutron monitor response\citep{2015AdSpR..55..363M}. Specifically, We have $\phi_{\rm IMP}$ = 0.666 GV, $\phi_{\rm Ulysses}$ = 0.7266 GV, $\phi_{\rm ACE}$ = 0.6093 GV, $\phi_{\rm ISEE}$ = 0.7409 GV, $\phi_{\rm ISOMAX}$ = 0.5974 GV, $\phi_{\rm PAMELA}$ = 0.5738 GV.
%\textcolor{red}{Still not clear how it was calculated??? In Zhu et al., data from 1997 to 2017 were given. It means only for ACE and ISOMAX the modulation parameter can be obtained? Due to the lack of measurement data of $\rm ^{10}Be/^{9}Be$ during the same period as AMS-02, the method mentioned in reference \citep{2018ApJ...863..119Z} was used to calculate the modulation potential, and the solar modulation of $\rm ^{10}Be/^{9}Be$ was estimated to be $\phi_{\rm ^{10}Be}/\phi_{\rm ^{9}Be}=0.49$.}
%The data are summarized in Table \ref{table:data}.

\begin{table}
\begin{center}
\caption[]{Data used in the fitting.}
\begin{tabular}{ccccc}
\hline\noalign{\smallskip}
 & Experiment & Time & Modulation & Ref.\\
\hline\noalign{\smallskip}
B \& C \& O & Voyager  & 2012/12-2015/06 &$\phi$& \citet{2016ApJ...831...18C} \\
    & ACE     & 2011/05-2018/05 &$\phi$& \citet{2019SCPMA..6249511Y} \\
    & AMS-02  & 2011/05-2018/05 &$\phi$& \citet{2021PhR...894....1A} \\
\hline
$^{10}$Be/$^9$Be & IMP     & 1974/01-1980/05 & 0.67 GV & \citet{1988SSRv...46..205S} \\
                 & Voyager & 1977/01-1998/12 & 0.78 GV & \citet{1999ICRC....3...41L} \\    
                 & ISEE    & 1978/08-1979/08 & 0.74 GV & \citet{1980ApJ...239L.139W} \\
                 & Ulysses & 1990/10-1997/12 & 0.73 GV & \citet{1998ApJ...501L..59C} \\
                 & ACE     & 1997/08-1999/04 & 0.61 GV & \citet{2001ApJ...563..768Y} \\
                 & ISOMAX  & 1998/08-1998/08 & 0.60 GV & \citet{2004ApJ...611..892H} \\
                 & PAMELA  & 2006/07-2014/09 & 0.57 GV & \citet{2021Univ....7..183N}\\
\noalign{\smallskip}\hline
\end{tabular}
\label{table:data}
\end{center}
\end{table}

Our MCMC process involves 14 parameters, including 5 propagation parameters 
($D_0$, $\delta$, $\eta$, $z_h$, and $v_A$), 6 injection spectral parameters, 
2 normalization parameters for carbon and oxygen, and 1 parameter for solar 
modulation. The CR nuclei of $Z\leq14$ were included, with normalization 
parameters being set to the default values of GALPROP except carbon and
oxygen. 

%%%%%%%%During this process, we make minor updates to the GALPROP code and adjust the $dz$ parameter to ensure the earth's Z-direction location is always at $Z=0$. This small change resolves the challenges of parameter degeneracies and improves the acceptance of MCMC sampler~\cite{2020JCAP...11..027Y}. As a result, the errors of parameters and the fluxes of GALPROP results are more precise. Without this modification, the parameters $zh$ may be limited to a small range and may not converge.

\begin{figure*}[htbp]
\begin{center}
\includegraphics[width=0.49\textwidth]{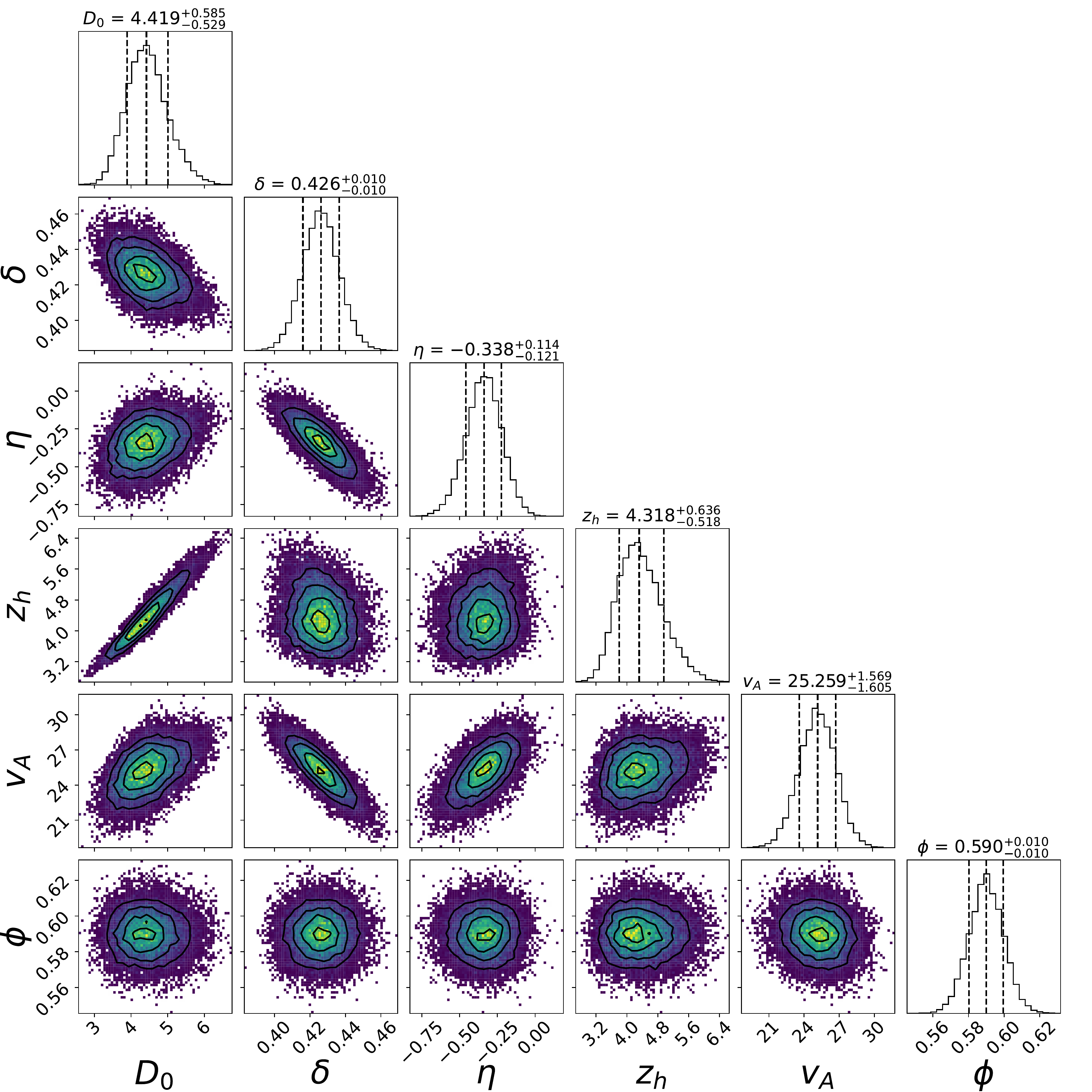}
\includegraphics[width=0.49\textwidth]{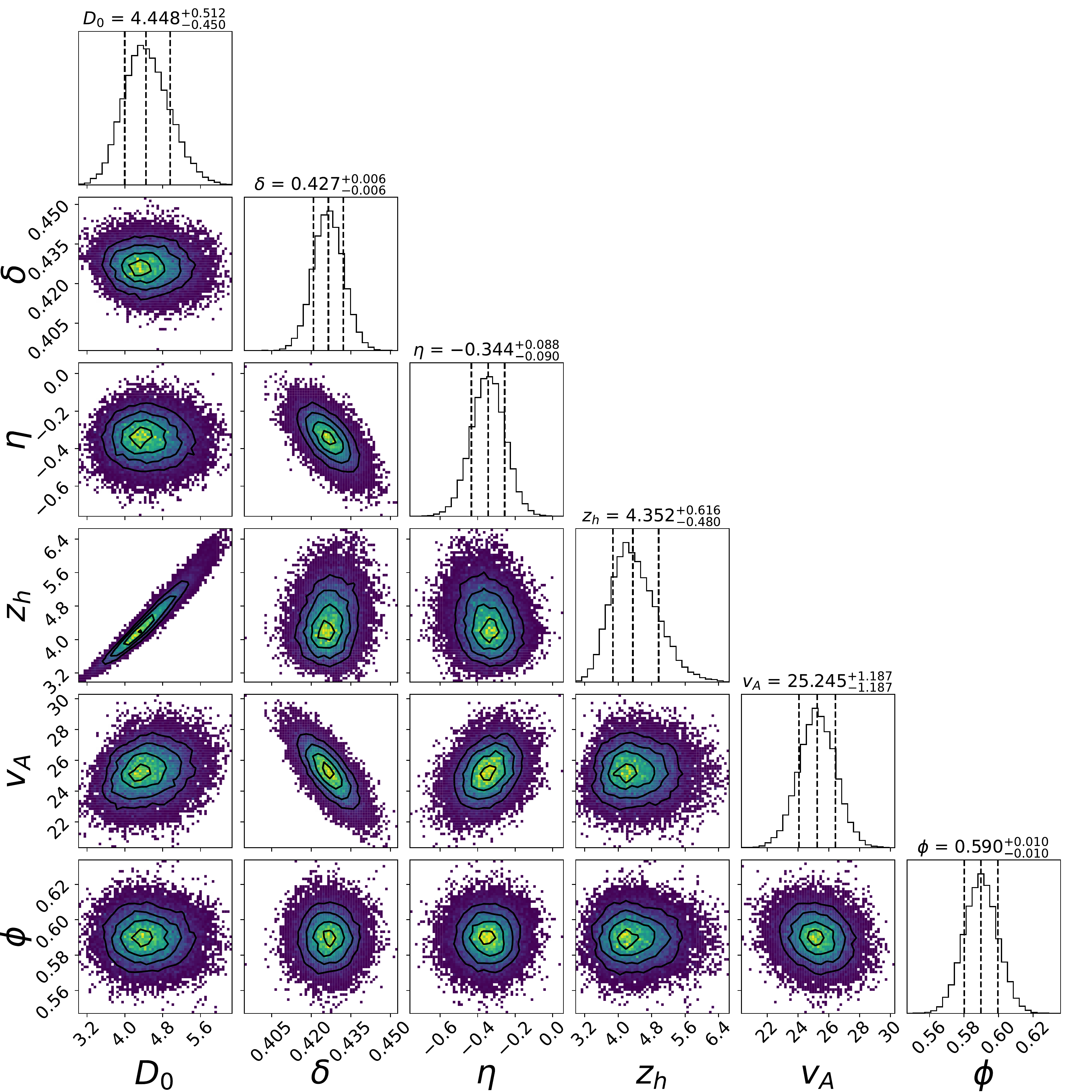}
\caption{The one-dimensional (diagonal) and two-dimensional (off-diagonal) probability 
distributions of the propagation parameters. For units of the parameters please refer 
to Table~\ref{table:prop}. The left panel is for the fitting to the existing data and 
the right panel is for the fitting to existing data plus the HERD predicted spectra.}
\label{fig:emcee_BC}
\end{center}
\end{figure*}

The posterior mean and $1\sigma$ uncertainties of the propagation parameters
and modulation parameter of the fitting are given in Table~\ref{table:prop}. 
The one-dimensional distributions and their two-dimensional correlations are shown 
in the left panel of Fig.~\ref{fig:emcee_BC}. The best-fitting spectra of boron, 
carbon, and oxygen, compared with the data, are shown in Fig.~\ref{fig:BC_Pre}.
We note that the derived model parameters are slightly different from those 
obtained previously, e.g., \citet{2020JCAP...11..027Y}. 
Specifically, the thickness of the propagation halo is smaller, the Alfven speed 
is smaller, and the parameter $\delta$ is bigger in this work compared with 
\citet{2020JCAP...11..027Y}. There are several possible reasons of such differences. 
Firstly, the data of AMS-02 used in this work are the seven-year measurements 
\citep{2021PhR...894....1A}. Secondly, we use version 56 of GALPROP in this work, 
which includes an update of the gas model and results in slightly different 
production spectra of secondary particles. Thirdly, different data sets used 
in these works may also result in different constraints on the model parameters. 
Our results are closer to those given in a recent work \citep{2023PhRvD.107f3020Z}, 
where some of the above mentioned updates have been included.

\begin{table}[!htb]
\begin{center}
\caption[]{Fitting results and $1\sigma$ uncertainties of the transport and solar modulation
parameters. Two fittings without and with the HERD data are carried out for comparison.}
\setlength{\tabcolsep}{7mm}{
\begin{tabular}{cccccccc}
\hline\noalign{\smallskip}
Parameter                  &  Fit without HERD  & Fit with HERD \\

\hline\noalign{\smallskip}
$D_0$ ($10^{28}$~cm$^2$~s$^{-1}$) & $4.42^{+0.59}_{-0.53}$   & $4.45^{+0.51}_{-0.45}$  \\
   $\delta$                    & $0.426^{+0.010}_{-0.010}$ & $0.427^{+0.006}_{-0.006}$  \\
   $\eta$                      & $-0.34^{+0.11}_{-0.12}$  & $-0.34^{+0.09}_{-0.09}$ \\
   $z_h$ (kpc)                 & $4.32^{+0.64}_{-0.52}$  & $4.35^{+0.62}_{-0.48}$  \\
   $v_A$ (km s$^{-1}$)         & $25.26^{+1.57}_{-1.61}$  & $25.25^{+1.19}_{-1.19}$ \\
%   $\chi^2_{\rm min}/dof$      & $135.8/254$  & $136.0/273$  \\ 
\hline
   $\phi$ (GV)                 & $0.59^{+0.01}_{-0.01}$ & $0.59^{+0.01}_{-0.01}$  \\
\noalign{\smallskip}\hline

%\hline\noalign{\smallskip}
%$D_0$ ($10^{28}$~cm$^2$~s$^{-1}$) & $3.64^{+0.50}_{-0.46}$   & $3.62^{+0.45}_{-0.40}$  \\
%   $\delta$                    & $0.426^{+0.010}_{-0.010}$ & $0.427^{+0.006}_{-0.006}$  \\
%   $\eta$                      & $-0.34^{+0.11}_{-0.12}$  & $-0.34^{+0.09}_{-0.09}$ \\
%   $z_h$ (kpc)                 & $3.49^{+0.47}_{-0.42}$  & $3.48^{+0.45}_{-0.38}$  \\
%   $v_A$ (km s$^{-1}$)         & $25.10^{+1.57}_{-1.55}$  & $25.08^{+1.22}_{-1.23}$ \\
%   $\chi^2_{\rm min}/dof$      & $135.8/254$  & $136.0/273$  \\ 
%\hline
%   $\phi$ (GV)                 & $0.590^{+0.010}_{-0.010}$ & $0.589^{+0.010}_{-0.009}$  \\
%\noalign{\smallskip}\hline
\end{tabular}}
\label{table:prop}
\end{center}
\end{table}

\subsection{Predicted boron, carbon and oxygen spectra by HERD}

\begin{figure*}[htbp]
\begin{center}
\includegraphics[width=\textwidth]{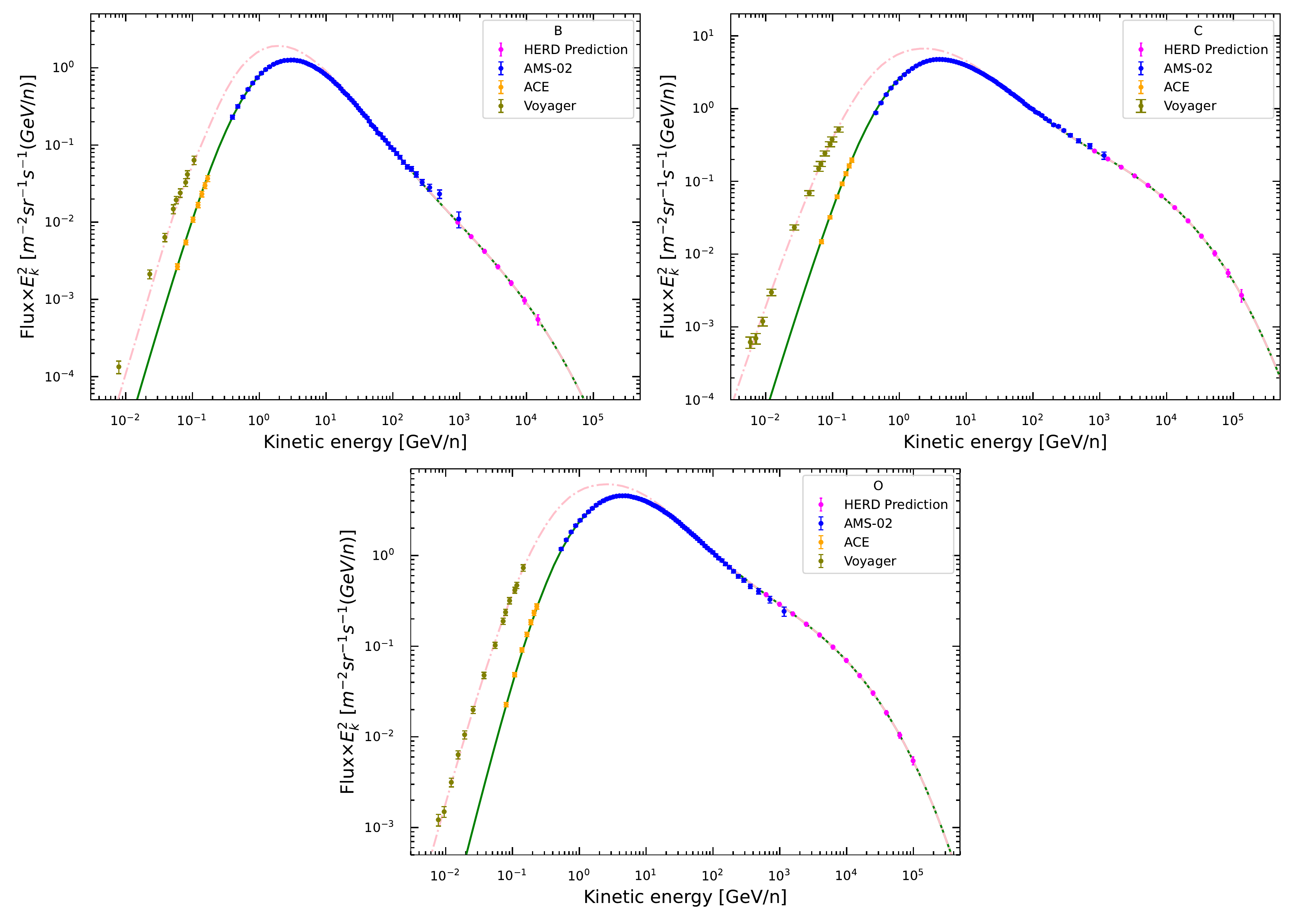}
\caption{The boron (top-left), carbon (top-right) oxygen (bottom) spectra. Lines are the best-fitting results
obtained in Sec. 3.1, with solid (dashed) ones being the spectra after (before) the solar
modulation. The HERD predicted data points are shown by pink dots.}
\label{fig:BC_Pre}
\end{center}
\end{figure*}

Based on the best-fitting spectra of CR boron, carbon, and oxygen, we forecast the prediction of 
HERD measurements. We choose a bin width of $\Delta \log E = 0.2$ of energies, and calculate the 
expected number of counts in each energy bin based on the simulated effective acceptance of 
the latest HERD design, for an operation time of 10 years. The effective acceptance takes into 
account both the geometry factor and the shower development in the CALO based on Monte Carlo 
simulations. Only the events with early developing showers with sufficient path lengths in 
the CALO are selected to ensure a good reconstruction quality. To enable a good flux measurement, 
we further require that the number of events in each energy bin is bigger than 10. 
Statistical uncertainties and the estimated systematic uncertainties of $\sim10\%$ 
\citep{2019SciA....5.3793A,2021PhRvL.126t1102A} are added in quadrature. 
The predicted fluxes measured by HERD, together with other experimental data are shown in
Fig.~\ref{fig:BC_Pre}. 

%and other experimental data, as well as the best-fit results obtained in the previous process, are shown in Fig.~\ref{fig:BC_Pre}. For the predicted boron and carbon fluxes, the last data point has counts greater than 10. The model decreases above 10 TeV/n due to the uncertainty of injection parameters at high energy. As a result, the statistical error of the predicted fluxes in this energy range increases. Nevertheless, our conclusion can still provide a relatively strong constraint on the transport parameters. If the predicted fluxes are higher, our constraints will be even more robust~\citep{2019A&A...627A.158D,2020PhRvR...2d3017H}.

\begin{figure*}[htbp]
\begin{center}
\includegraphics[width=0.6\textwidth]{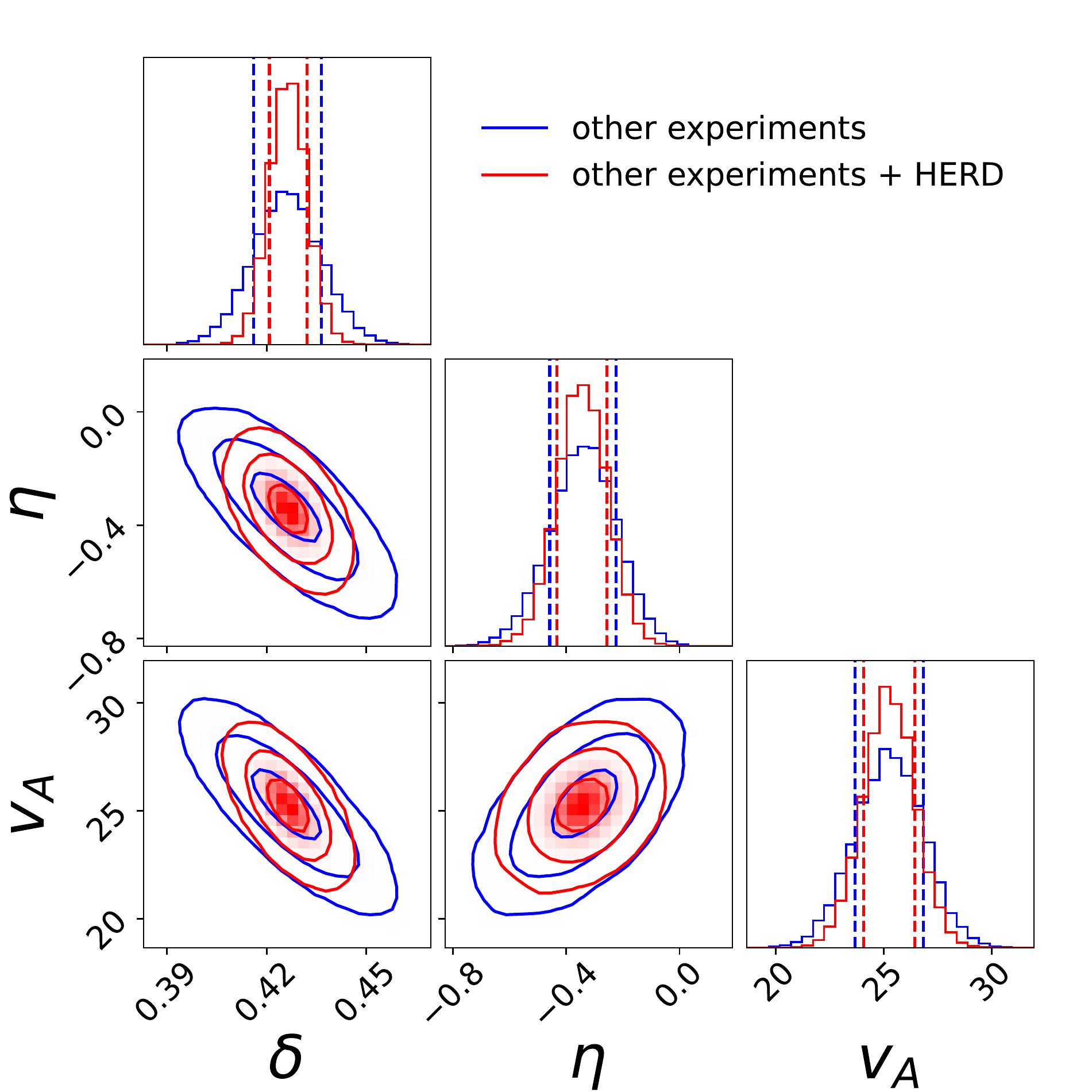}
\caption{Comparison of the constraints on selected propagation parameters for the fitting
without (blue) and with (red) HERD data.}
\label{fig:emcee_overlap}
\end{center}
\end{figure*} 

%-----------------------------------------------------------------

%-------------------------------------- Two column figure (place early!)
\subsection{Constraints on model parameters by including HERD}
We re-do the fitting as done in Sec. 3.1 to derive the constraints on the model parameters
after including the HERD spectra on boron, carbon, and oxygen. The results are given in 
Table~\ref{table:prop} and Fig.~\ref{fig:emcee_BC}. We find that adding the HERD data will 
reduce the errors of model parameters by about 8\% to 40\%. The improvement on the constraint 
of $\delta$ is the most significant. This is expected since HERD mainly improves the measurements 
at high energies. Note that new measurements of AMS-02 and DAMPE showed hardenings of the B/C and 
B/O ratios at high energies \citep{2021PhR...894....1A,2022SciBu..67.2162D}, which are not 
included in this work. However, it is expected that HERD can definitely give better measurements 
of such ratios above TeV/n, and can thus better constrain the energy dependence of the diffusion
coefficient at high energies. There are slight improvements on other propagation parameters,
which are mainly determined by low-energy data. Fig.~\ref{fig:emcee_overlap} compares the
constraints on parameters $\delta$, $\eta$, and $v_A$, for the fittings without (blue) and 
with (red) HERD data.

\begin{figure*}[htbp]
\begin{center}
\includegraphics[width=0.6\textwidth]{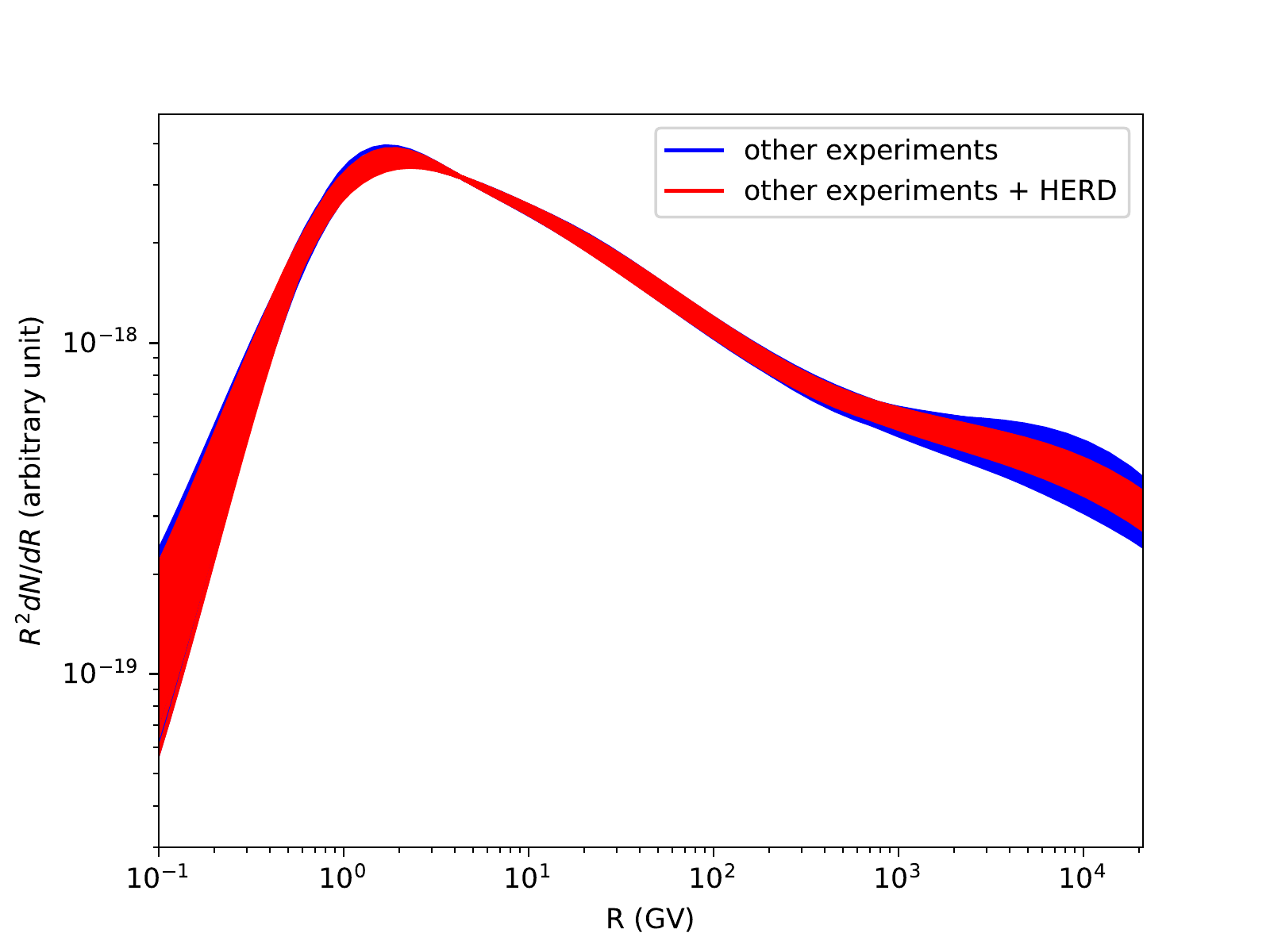}
\caption{The $1\sigma$ bands of injection spectra for the fitting without (blue) and
with (red) HERD data.}
\label{fig:In_sp}
\end{center}
\end{figure*}

The inclusion of HERD data at high energies is expected to improve the constraints on the
wide-band injection spectra of CRs. Fig.~\ref{fig:In_sp} shows the comparison of the injection 
spectra of the two fittings. The shaded bands represent the $1\sigma$ spans of the fitting
results. As can be seen, the uncertainties above TV are smaller when adding the HERD data.

\section{Summary and discussion}
\label{sec:conclusion}

Precise measurements of spectra of primary and secondary CRs in a wide energy range are
very important in probing the propagation of Galactic CRs. In this work we study the perspective
of constraining CR propagation parameters with the planned HERD mission onboard China's Space
Station. HERD is expected to measure precisely energy spectra of carbon and oxygen nuclei up to 
100 TeV/n and boron nuclei to $>10$ TeV/n. These measurements are expected to be very helpful in 
understanding the energy dependence of the diffusion coefficient as well as the injection
spectra above TeV/n which are rarely constrained by existing data.

Under a framework of continuous source distribution, spatially homogeneous propagation with
reacceleration, we fit the boron, carbon, and oxygen data to obtain the constraints on the
model parameters. We focus on the comparison of the results for the fittings without and with
the HERD data. It is shown that adding the HERD data the constraint on the slope parameter
($\delta$) of the energy dependence of the diffusion coefficient gets improved significantly. 
Quantitatively the error of $\delta$ decreases by about $40\%$ after adding the HERD data.
Slight improvements on other propagation parameters are also found. In addition, the HERD
data are useful in improving the constraints on the injection spectra of primary CRs at high
energies.

The model assumption of the current work is simplified. Possible improvements of future works
may include 1) study of more secondary (such as lithium, beryllium, fluorine, sub-iron) and 
primary (oxygen, neon, magnesium, silicon, iron) CR spectra by HERD and particularly the effects 
of different mass groups \citep{2019PhLB..789..292W,2022arXiv220801337F,2023PhRvD.107f3020Z}, 
2) discussion of spectral breaks of the diffusion coefficient 
\citep{2012ApJ...752...68V,2023FrPhy..1844301M}, 
and 3) investigation of spatially inhomogeneous propagation as indicated by recent observations \citep{2012ApJ...752L..13T,2018PhRvD..97f3008G,2021PhRvD.104l3001Z}.

%%%%%%%%%%%%%%%%%%%%%%%%%%%%%%%%%%%%%%%%%%%%%%%%%%%%%%%%%%%%%%%%%%%%%%
\begin{acknowledgements}
We thank Wei Jiang, Zhao-Qiang Shen, and Cheng-Rui Zhu for helpful discussion.
This work is supported by the National Key Research and Development Program of China 
(No. 2021YFA0718404), the National Natural Science Foundation of China (No. 12220101003), 
and the Project for Young Scientists in Basic Research of Chinese Academy of Sciences 
(No. YSBR-061). 
\end{acknowledgements}

\bibliographystyle{raa}
\bibliography{ref1}

\label{lastpage}

\end{document}